\documentclass[10pt,journal,final]{IEEEtran}
\usepackage{multirow}
\usepackage{booktabs}
\usepackage{makecell}
\usepackage{amsmath,bm}
\usepackage{cite}
\usepackage{graphicx}
\usepackage{epstopdf}
\graphicspath{{./figures/}}
\usepackage{amsmath}
\usepackage{amssymb}
\usepackage[ruled]{algorithm2e}
\usepackage{caption}
\usepackage{amssymb} 
\usepackage{subfigure}
\usepackage{stfloats}
\usepackage{enumitem}
\newcommand{\RNum}[1]{\uppercase\expandafter{\romannumeral #1\relax}}
\let\sss= \scriptscriptstyle
\begin{document}
	\title{Delay Aware Secure Offloading for NOMA-Assisted Mobile Edge Computing in Internet of Vehicles
	}
	\author{ \quad
		Ling He,~\IEEEmembership{Student Member,~IEEE},~Miaowen Wen,~\IEEEmembership{Senior Member,~IEEE},\\~Yingyang Chen,~\IEEEmembership{Member,~IEEE},~Bingli Jiao,~\IEEEmembership{Senior Member,~IEEE}
	\thanks{L. He and M. Wen are with the School of Electronic and Information Engineering, South China University of Technology, Guangzhou 510640, China (e-mail: eeheling@mail.scut.edu.cn; eemwwen@scut.edu.cn).}   
	\thanks{Y. Chen is with the College of Information Science and Technology, Jinan University, Guangzhou 510632, China (e-mail: chenyy@jnu.edu.cn).}
	\thanks{B. Jiao is with the Department of Electronics, Peking University, Beijing 100871, China (e-mail: jiaobl@pku.edu.cn).}
}	
	\maketitle
	\begin{abstract} 
    In this paper, a multi-vehicle multi-task non-orthogonal multiple access (NOMA) assisted mobile edge computing (MEC) system with passive eavesdropping vehicles is investigated. To heighten the performance of edge vehicles, we propose a vehicle grouping pairing method, which utilizes vehicles near the MEC as full-duplex relays to assist edge vehicles. For promoting transmission security, we employ artificial noise to interrupt eavesdropping vehicles. Furthermore, we derive the approximate expression of secrecy outage probability of the system. The combined optimization of vehicle task division, power allocation, and transmit beamforming is formulated to minimize the total delay of task completion of edge vehicles. Then, we design a power allocation and task scheduling algorithm based on genetic algorithm to solve the mixed-integer non-linear programming problem. Numerical results demonstrate the superiority of our proposed scheme in terms of system security and transmission delay. 
	\end{abstract}

	
	\IEEEpeerreviewmaketitle
	
	\section{Introduction}
	\subsection{Motivation and Scope}
   Recently, the Internet of Vehicles (IoV) has attracted increasing attention from both academics and industry. Notably, the number of registered motor vehicles is more than 1 billion until now and is expected to double within the next 10-20 years \cite{b1}. This impels the IoV to become one of the major trends in the automotive industry field gradually. With the continuous development of wireless communication technology, the application of IoV is not only expected to realize lower delay and higher reliability communication \cite{b2}, but also to provide more advanced and complex services, such as automatic driving, vehicle formation driving, and other computationally intensive and time-delay sensitive services in 5G \cite{b3}. Unfortunately, the vehicle cannot fulfill the high quality and low latency requirements due to its own computing capability and limited battery life \cite{b4}. 
	 
	 In order to solve this problem, mobile edge computing (MEC) has become a feasible solution that serves as an emerging data processing technology. By offloading tasks from the vehicles to a nearby MEC server, the computing capability of the vehicle can be significantly enhanced \cite{b5}. Generally, there are two task offloading modes in MEC, which are binary computing offloading \cite{b7} and partial computing offloading \cite{b9}. In the binary computing offloading mode, computing tasks cannot be partitioned. The task is either executed locally on the mobile device or completely offloaded to the MEC server. In the partial computing offloading mode, the calculation task can be divided into two parts: one part is executed locally, and the other part is offloaded to the MEC server.
	 
	 There are several works considering the combination of orthogonal multiple access (OMA) and MEC \cite{b10,b11,b12}. However, the per capita car ownership rate has increased rapidly in recent years. Meantime, recent research found that non-orthogonal multiple access (NOMA) technology has the advantages of improved spectrum efficiency and strengthened robustness in high mobility schemes \cite{b15}. Moreover, Ding \textit{et al.} demonstrated that NOMA can effectively reduce energy consumption and offload delay in MEC networks compared with OMA \cite{b17}. Therefore, to meet the demand of large-scale connectivity in IoV, NOMA technology has been regarded as a potential replacement of OMA \cite{b19}. 
	 
	 In addition, data security and privacy protection issues are crucial for the NOMA-MEC system in IoV. Specifically, due to the broadcast characteristics of the cellular wireless communication channel, it is very easy for the eavesdroppers to obtain task data offloaded by a legitimate vehicle to the MEC server. Previous research mostly used encryption technology to protect data \cite{b20}. Recently, physical layer security (PLS) has emerged as an effective technology to exploit the different dynamic characteristics between a legal wireless channel and an eavesdropper's channel, to improve the communication security \cite{b21,b22}. Compared with the high implementation complexity of encryption technology, PLS does not require additional costs, which has been regarded as a powerful complementary solution for enhancing the security of cellular wireless communication systems.
	
	\subsection{Related Works}
	 Researches on the resource optimization of the NOMA-MEC system can be divided into three categories: minimizing energy consumption \cite{b23,b24,b25}, minimizing time delay \cite{b28,b26,b27} and minimizing system cost \cite{b30,b29}.
	
	In order to reduce energy consumption in the edge computing environment, Kiani \textit{et al.} proposed a heuristic user clustering scheme based on the order of user channel gains, which minimizes the total energy consumption of users by jointly optimizing user allocation, computing resource allocation, and transmission power allocation\cite{b23}. The authors of \cite{b24} utilized NOMA in the MEC system for task offloaded and result download, then transmit power control, transmission time allocation, and task offloading partitions were jointly optimized to minimize the total energy consumption. Moreover, in \cite{b25}, a NOMA-MEC scheme that uses an assistant to assist users in data offloading was proposed, which minimizes the energy consumption and maximizes the data offloaded by jointly optimizing the allocation of communication and computing resources between users and assistants.
	
	Besides energy consumption, the minimization of latency is also a hot issue in terms of the ultra-low latency requirement of 5G communication. The study of \cite{b26} jointly optimized user unloading workload and transmission time to minimize the total delay of users. In \cite{b27}, the successive interference cancellation (SIC) sequencing and computing resources of the NOMA-MEC network were optimized to minimize the maximum task execution delay. Furthermore, Sheng \textit{et al.} described the interaction between the differential upload delay of a pair of NOMA users and co-channel interference \cite{b28}. Besides, the authors provided a scheme to reduce the average offloading delay of users by jointly optimizing the offloading decision and resource allocation.   
	
	There exist some works studying the minimization problem of system cost, which comprises the weighted sum of time and energy consumption. For example, Yang \textit{et al.} minimized the linear combination of the completion time of all tasks and the total energy consumption (including transmission energy and local computing energy) when the upload data rate and edge cloud computing capabilities are limited \cite{b29}. The authors of \cite{b30} realized the weighted value minimization of time and energy consumption by jointly optimizing the power and the time allocation between each group and user groups.
	
    On the other hand, PLS has received extensive attention to protect the data security of legitimate users in the literature \cite{b31,b32,b33,b34,b35,b36,b37}. The authors of \cite{b31} realized the safe offloading of computing tasks through joint optimization of security offloading and partial computing. Furthermore, an energy-saving computing offloading problem was formulated in the presence of active and passive eavesdroppers in \cite{b32}. Besides, the study of \cite{b33}  investigated the design of NOMA against the external eavesdropper given the secrecy outage constraint, to simplify the confidentiality capacity analysis, where the perfect channel estimation is assumed at the eavesdropper. However, the channel state information (CSI) acquisition problem is quite difficult in actual passive eavesdropping scenarios. For that reason, the authors of \cite{b34} studied the problem of maximizing the minimum value of the confidential information rate under the imperfect instantaneous CSI scenario. Particularly, an artificial noise (AN) scheme was used to interfere with potential eavesdroppers for further improving the communication security in \cite{b36, b37}.
	
	\subsection{Contributions and Organization}
    Most of the above researches on NOMA-MEC network \cite{b23,b24,b25,b26,b27,b28,b29,b30} concentrates on two scenarios in IoV: 1) each vehicle has only one task; 2) only two vehicles exploit NOMA to offload task. However, to meet the demands of multiple business scenarios in IoV, handling multiple tasks, such as intelligent navigation and entertainment information services tasks, is needed in the meantime. Besides, the above-mentioned researches pay less attention to security issues in the NOMA-MEC network, especially in an IoV scenario. Motivated by these issues, our research focuses on a multi-vehicle multi-task offloading process with a single MEC in the traffic congestion scenario. The main purpose is to minimize the total completion time of the vehicle tasks' secure offloading.
    
	In our NOMA-MEC network, a passive eavesdropping vehicle randomly distributes. Since the channel of eavesdropper is unknown, it is challenging to realize the secure communication of legitimate vehicles through channel differences. Therefore, we use base station (BS) to send AN signals for interfering with potential eavesdroppers. In order to ensure a secure and low-latency completion of the tasks of the edge vehicles, the central vehicles are utilized as a full-duplex (FD) relay \cite{b38} to assist the edge vehicles for data transmission, thus ensuring the low-latency completion of the edge vehicles' tasks. The main contributions of this paper are summarized as follows:  
	\setlist{nolistsep}
    \begin{itemize}
    \item We investigate a multi-vehicle multi-task NOMA-MEC system with a passive eavesdropping vehicle. In order to improve the quality of service of edge vehicles, we propose a multi-vehicle grouping and pairing method (GPM) for dividing the randomly distributed vehicles into two groups, i.e., the center group and the edge group, then matching a pair of vehicles one from each group, where the vehicle from center group is used as an FD relay to assist the edge vehicle. 
   
   \item  To improve the data security, the BS is used for transmitting AN to interfere with the eavesdropping vehicle, and by utilizing the null-steering beamforming technology to set the antenna transmission weights, the interference caused by AN to legitimate vehicles can be avoided. Furthermore, based on GPM and AN, for a pair of vehicles, we derive the exact and asymptotic expressions for the secrecy outage probability (SOP) of the center and edge vehicles, and the approximate expression of the SOP of the system (SOPS).

	\item To minimize the total delay of task completion of edge vehicles, we design a power allocation and task scheduling algorithm relying on genetic algorithm (GA-PATS) to jointly optimize the vehicle task division, power allocation and transmit beamforming. The excellent performance of the proposed scheme in terms of system security and transmission delay is proved by simulations.   
   \end{itemize}

	The organization of this paper is as follows. The system model is introduced in Section II, where the GPM is proposed. In Section III, the SOPS for the NOMA-MEC network is derived. The task offloading time minimization problem is solved in Section IV. Simulation results are presented in Section V. Section VI concludes the paper.	
	
    \textit{Notation}: Capital bold and lowercase letters represent matrices and vectors, respectively. $ (\cdot)^{T} $ and $ (\cdot)^{H} $ stand for transpose and Hermitian transpose, respectively. $ x \sim \mathcal{CN}(a,b) $ means that the scalar \textit{x} follows a complex Gaussian distribution with mean \textit{a} and covariance \textit{b}. $ x \sim \mathcal{N}(0,\sigma^2) $ means that the scalar $ x $ follows a Gaussian distribution with with zero mean and variance $ \sigma^2 $. $ \left\|\cdot\right\| $ is the Euclidean norm. $  F(\cdot) $ and $ f (\cdot) $ are the cumulative density function (CDF) and probability density function (PDF), respectively. $ \mathrm{E}_i(\cdot) $ is the exponential integral function, that is, $ \mathrm{E}_i(a, b)=\int_{1}^\infty e^{-xb}x^{-a}dx $. $ P\left\{X < x\right\} $ is the probability that the random variable $ X $ is less than $ x $. Finally, $ \mathbf{H} \in \mathbb{C}^{M\times N} $ indicates that $ \mathbf{H} $ is a complex-element matrix with dimensions $ {M\times N} $.

	\setlength{\textfloatsep}{5pt}
    \begin{figure*}[htp] 
	\centering 
	\includegraphics[scale=0.81]{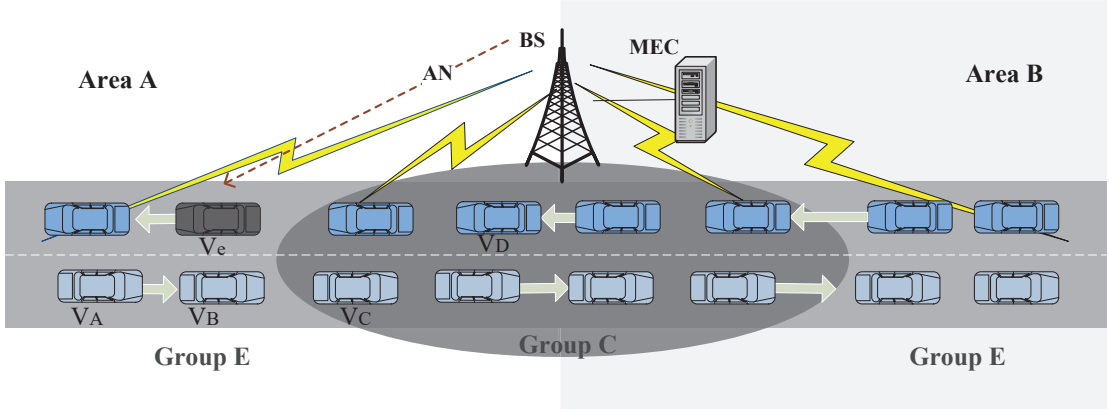}	
	\caption{Single MEC multi-vehicle grouping system model}
	\label{fig1}
    \end{figure*} 

	\section{System Mode}
	As shown in Fig. \ref{fig1}, we consider a multi-vehicle multi-task NOMA-MEC network with a passive eavesdropping vehicle. In this network, the BS and the MEC server are connected by optical fiber, which can be regarded as a whole. The BS equipped with $\mathit{K}$ antennas is located in the cell center, whose working mode is FD. A two-way lane with $\mathit{N}$ legitimate vehicles and one eavesdropping vehicle randomly distributed in the coverage of BS, and the direction close to the MEC is defined as the positive direction. Additionally, each legal vehicle is equipped with an FD antenna and has $\mathit{M}$ tasks to be completed at the same time. All channels follow quasi-static Rayleigh fading. Without loss of generality, we assume that all the  additive white Gaussian noises (AWGN) are independent and identically distributed (i.i.d) with a distribution ${\mathcal{CN}(0,\sigma^{2}_n)}$.

	\subsection{Vehicle Grouping and Pairing}
	The core idea of NOMA is to allocate more power to users with worse channel conditions \cite{b39}. In particular, these users with better channel conditions know the message intended for the others by applying SIC and thus can act as relays to improve reception reliability and transmission rates for users with worse channel conditions \cite{b40}. On the other hand, even if SIC is adopted, the problem of co-channel interference still exists. Moreover, the more users sharing the same resource block, the stronger the co-channel interference meets, so two users are a pair generally. 
		
	Based on this, we propose a multi-vehicle grouping and pairing method (GPM). Suppose that the coverage area of MEC is a circular area with a radius of \textit{R}. Furthermore, BS can obtain location information of all legal vehicles. According to the straight-line distance between the legal vehicles and MEC, all legal vehicles are divided into two groups, namely the center group (group C) and the edge group (group E). Specifically, these vehicles within the radius of $R_{MC}$  belong to group C, other legal vehicles belong to group E, and the eavesdropping vehicles are randomly distributed in the range of 0 to \textit{R}.	
			
	Before vehicle pairing, the coverage area of BS is divided into two areas A and B. Here, only the pairing method of area A is discussed. The pairing method of area B is the same as that of area A. It is reasonable to ignore the case where one vehicle from area A and another from area B are taken as a pair because the distance between them is relatively long and the transmission channel condition is poor.
	
	As shown in Fig. \ref{fig1}, we denote the direction close to the MEC as the positive direction. The main objective in considering vehicle pairings is to provide as much time as possible for edge vehicles when the central vehicle acts as a relay. Considering that the farther the vehicle is from MEC, the worse the communication channel conditions from the vehicle to MEC and the lower the transmission rate will be, so we give priority to the farthest vehicle from MEC in group E. Our proposed GPM is summarized as follows:
	\begin{itemize}
		\item When the edge vehicle is moving towards MEC and there exist vehicles moving in the positive direction in group C, the nearest vehicle in group C is selected according to the principle of physical proximity. In this way, we can determine that  ${\mathrm{V}_A}$ and ${\mathrm{V}_C}$ are in a group in Fig. \ref{fig1}.
		\item When all the rest of vehicles in group C are moving in the opposite direction of the edge vehicle, according to the principle of the longest physical distance, the farthest vehicle in group C is selected, e.g.,  ${\mathrm{V}_B}$ and ${\mathrm{V}_D}$ form a pair in Fig. \ref{fig1}. 
		\item When the edge vehicle is traveling in negative direction, then the vehicle pair is selected according to the principle of taking the furthest vehicle in negative direction and the nearest in positive direction.
	\end{itemize} 
	
     The details for GPM can be found in Algorithm 1. In the following some parameters in the GPM are described. The horizontal distance from each legitimate vehicle to MEC is ${l_n}$, $ n \in \left\{ 1,\cdots ,\mathit{N} \right\}$. Besides, the traveling speed of vehicle ${\nu_n}$, $ n \in \left\{ 1,\cdots ,\mathit{N} \right\}$ obeys the uniform distribution on the interval $\left[-\nu,\nu\right]$, and $ \nu $ (in m/s) is the maximum travelling speed. 	

\setlength{\intextsep}{4pt}
    \IncMargin{1em}
    \begin{algorithm}
	\LinesNumbered % Display the line number
	\caption{GPM Algorithm} 
	\label{GPM Algorithm}
	\KwIn{ \textit{N}, \textit{R}, $R_{MC}$, \textit{VD}, $ \nu $ }
	\KwOut{Vehicle pairing result \textbf{GP}}
	Generate $ V_n = [ \nu_n, l_n ], n= \left\{ 1,\cdots ,\mathit{N} \right\} $ and then get \textbf{V} = $ [ \textbf{v}, \textbf{l} ] $, where $ \textbf{v} = [\nu_1,\cdots ,\nu_N] $ and $ \textbf{l} = [l_1,\cdots ,l_N] $.\\
	According to the \textbf{l} and $R_{MC}$, \textbf{V} is divided into \textbf{C} and \textbf{E}.\\
	\For{$j = size(\textbf{E},1) : -1 : 1$}{
		\uIf{There are vehicles in the same  direction in \textbf{C}}{
			Select the nearest vehicle as a pair P; \\
		}\ElseIf{There are vehicles in the opposite  direction in \textbf{C}}{
		    Select the furthest vehicle as a pair P;\\
	    }
       Update \textbf{GP} = [ \textbf{GP} ; P ];\\
       Delete the vehicle selected in \textbf{C}.
	}
    \end{algorithm}
    \DecMargin{1em}
	\subsection{Task Offloading}
   In this paper, we adopt the hybrid offloading mode that was chosen  according to the transmission power. When the transmitting power is sufficient to ensure that all tasks are offloaded to the MEC in less time than the minimum tasks need to be executed locally, all tasks will be offloaded to the MEC. If channel conditions do not guarantee safe offloading, the method of full local execution is used. In other cases, a partial computing offloading mode will be adopted. Due to the large coverage area of the BS, the communication channel between the edge vehicles and MEC is poor, so the direct link between them is assumed to be negligible in this paper. The center vehicle can communicate directly with the MEC.
	
   In order to simplify the expression, this paper only analyzes the offloading of the \textit{n}-th vehicle ${\mathrm{V}_n}$, $ n \in \left\{ 1,\cdots,N \right\}$, and the task division of other vehicles is the same. The \textit{M} tasks on ${\mathrm{V}_n}$ are reordered according to the data size from small to large. The former ${m_n}$ tasks apply local execution pattern, and the remaining ${M-m_n}$ tasks are offloaded to MEC.  
 
   We denote the \textit{j}-th task on ${\mathrm{V}_n}$ as ${T_{n,j}}= \left\{ s_{n,j},c_{n,j}\right\}$, $ j \in \left\{ 1,\cdots,M \right\}$, where $ s_{n,j}$ is the size of the data contained in ${T_{n,j}}$, and $c_{n,j}$ is the number of CPU cycles required. Let \textit{c} is the number of CPU cycles required per bit (\textit{c} is a constant), we can get $c_{n,j}=cs_{n,j}$. $f_{Local}$ (in cycles/bit) denotes the frequency of the local CPU and $D^{max}_n$ reprents the maximum tolerable delay for the completion of all tasks of ${\mathrm{V}_n}$, $ n \in \left\{ 1,\cdots,N \right\}$. The total local calculation time is	
	\begin{equation}
	D^{Local}_n=\sum_{j=1}^{m_n} \frac{c_{n,j}}{f_{Local}}=\sum_{j=1}^{m_n} \frac{cs_{n,j}}{f_{Local}}.	
	\end{equation}     
    To simplify the analysis, we assume $\forall {D^{max}_n=D}$, $ n \in \left\{ 1,\cdots,N \right\}$ and $\forall s_{n,j}=s$, $ j \in \left\{ 1,\cdots,M \right\}$.
   	\setlength{\textfloatsep}{5pt}
   \begin{figure}[tt]
   	\centering
   	\includegraphics[scale=0.8]{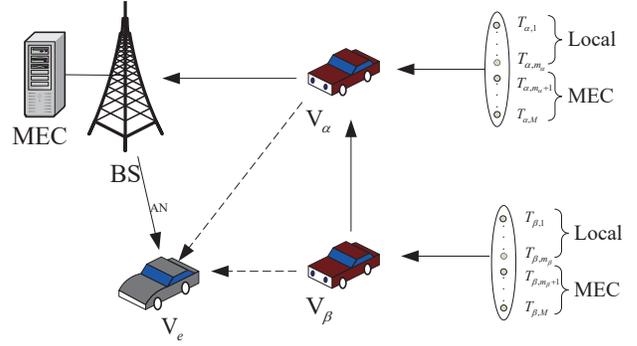}			
   	\caption{Central vehicle $ V_\alpha $ acts as an FD relay to assist edge vehicle  $ V_\beta $ in the presence of an eavesdropping vehicle $ V_e $ in the NOMA-MEC system}
   	\label{fig2}
   \end{figure} 
	\subsection{Communication Model}
	As we can see in Fig. \ref{fig2}, the vehicle in the center group is ${\mathrm{V}_\alpha}$, and the vehicle in the edge group is ${\mathrm{V}_\beta}$, where $\alpha,\beta \in \left\{ 1,\cdots,N \right\}  $. ${\mathrm{V}_e}$ is the eavesdropping vehicle whose instantaneous channel state information (CSI) is unknown. Their horizontal distances from MEC are $ l_\alpha $, $  l_\beta $, $ l_e $, respectively. The height of BS is $\mathit{L}$, and the road width is negligible.

   \subsubsection{AN}
   Since there is a passive eavesdropper with unknown CSI in the system, for ensuring the safe transmission of offloaded data, by utilizing the null-steering beamforming technology \cite{b41},  we generate AN at BS to disturb the eavesdropping vehicles meanwhile without affecting the BS and legal vehicles. One antenna on BS is used to receive signals, and the other $K-1$ antennas are used to transmit interference signals. 
   
	The channel coefficients of channels $\mathrm{BS}\rightarrow\mathrm{BS}$, $\mathrm{BS}\rightarrow{\mathrm{V}_\alpha}$, and $\mathrm{BS}\rightarrow{\mathrm{V}_e}$ are denoted by ${\mathbf{h}_{bb}} \in \mathbb{C}^{K-1 \times K-1}   $, ${\mathbf{h}_{b\alpha}} \in \mathbb{C}^{K-1 \times 1}$, and ${\mathbf{h}_{be}} \in \mathbb{C}^{K-1\times1}  $, respectively. Each element of ${\mathbf{h}_{bb}}$ obeys the distribution ${\mathcal{CN}(0, \sigma^{2}_{SI})}$, and each element of $\mathbf{h}_{b\alpha}$ and $\mathbf{h}_{be}$ obeys the distribution ${\mathcal{CN}(0, \sigma^{2}_i)}$, ${i\in({b\alpha},{be})}$, the AN generated by the BS is an $N\times1$ dimensional Gaussian vector, where each element is i.i.d with a distribution of ${\mathcal{N}(0, 1)}$.
	
	The index of the antenna receiving the signal on the BS can be obtained by ${\mathit{k}=\mathop{\arg\max}\limits_{i=1,2,\cdots,K} |\mathit{h}_{\mathit{\alpha b,i}}|^2}$, then the remaining antennas are used to transmit AN. Denote that the maximum power assigned by BS to transmit AN is $P_B$, and the transmitting power of each antenna is $w_k P_B$, $ \forall w_k\in(0,1) $, $ k=1,2,\cdots,K-1 $,  where $ w_k $ is the weight assigned to the \textit{k}-th transmitting antenna. Then the total signal transmitted by the remaining $K-1$ antenna is $ \mathbf{x}_{AN}=\mathbf{w} \sqrt{P_B}n_b$, where $ \mathbf{w} = [w_1,w_2,\cdots,w_{K-1}]^T $ is a $(K-1)\times1$ dimensional vector, and $ n_b $ is Gaussian white noise. In order not to interfere with BS and legal vehicles, $ \mathbf{w} $ needs to meet the nullification condition of $ \mathbf{w}^\mathit{H}\mathbf{h}_{bb}=0 $ and $ \mathbf{w}^H\mathbf{h}_{b\alpha}=0 $. 

   \subsubsection{Offloading}
   	The channel coefficients of channels $\mathrm{V}_\beta\rightarrow\mathrm{V}_\alpha$, $\mathrm{V}_\alpha\rightarrow\mathrm{BS}$, $\mathrm{V}_\beta\rightarrow{\mathrm{V}_e}$, and $\mathrm{V}_\alpha\rightarrow{\mathrm{V}_e}$ are $h_{\beta \alpha}$, $h_{\alpha b}$, $h_{\beta e}$ and $h_{\alpha e}$ respectively, where ${h_j\sim\mathcal{CN}(0, \sigma^{2}_j)}$, ${j\in({\beta \alpha},{\alpha b},{\beta e},{\alpha e})}$. The channel gain $ |h_j|^2 $ obeys the exponential distribution with parameter $ \gamma_j $, $ j\in(bb, b\alpha,be,\beta\alpha,\alpha b,\alpha e,\beta e) $. The effective channel gain is $\mathrm{g}_j=|h_j|^2 d^{-v}_j$, where $d_j$ is the straight-line distance between the two nodes, $v$ is the path loss coefficient, and the distance from each transmit antenna on the BS to ${\mathrm{V}_e}$ is regarded as the same.
   	
	Assuming the power of ${\mathrm{V}_\beta}$ is ${P_\beta}$,
	the signal sent by ${\mathrm{V}_\beta}$ is $ x_1=\sqrt{P_\beta}x_\beta $. When ${\mathrm{V}_\alpha}$ acting as an FD relay to receives the signal of ${\mathrm{V}_\beta}$, the influence of self-interference still remains after applying the self-interference cancellation technology. The signal received at  ${\mathrm{V}_\alpha}$ can be denoted as
	\begin{equation}
		y_\alpha=h_{\beta\alpha} d^{-\frac{v}{2}}_{\beta\alpha}x_1+h_{\alpha\alpha}x_1+n_\alpha,
	\end{equation}
	where ${\mathit{h}_{\alpha\alpha}\sim\mathcal{CN}(0, \sigma^{2}_{SI})}$ is the self-interference channel coefficient.
	
	Denote the power of  ${\mathrm{V}_\alpha}$ as ${P_\alpha}$,
	in which part of the ${P_\alpha}$ is used to transmit ${\mathrm{V}_\beta}$'s signal and the rest is used to send own signal. As a result, the superimposed signal to be sent by  ${\mathrm{V}_\alpha}$ can be written as
	\begin{equation}
		x_{\sss 2}=\sqrt{\lambda P_\alpha}x_\alpha+\sqrt{(1-\lambda)P_\alpha}x_\beta,
	\end{equation}
	where $ \lambda $ is the power allocation coefficient (PAR) and according to the principle of NOMA \cite{b42}, ${\lambda\in(0,0.5)}$ should be satised. 
	
	Then, the received signal at the BS and the eavesdropper are respectively  given by 
	\begin{equation}
	y_b=h_{\alpha b}d^{-\frac{v}{2}}_{\alpha b}x_2+n_b,
	\end{equation}
	\begin{equation}
	y_e=h_{\beta e}d^{-\frac{v}{2}}_{\beta e}x_1+h_{\alpha e}d^{-\frac{v}{2}}_{\alpha e}x_2+d^{-\frac{v}{2}}_{b e}{\mathbf{h}^H_{b e}}\mathbf{x}_{AN}+n_e,
	\end{equation}
   where $ \mathit{n}_\alpha $, $ \mathit{n}_b $ and $ \mathit{n}_e $ are AWGN.
	
	Suppose a decode-and-forward (DF) protocol is applied at the ${\mathrm{V}_\alpha}$. Specifically, ${\mathrm{V}_\alpha}$ decodes its received signal under the NOMA protocol and then re-encodes and forwards the signal of ${\mathrm{V}_\alpha}$ and ${\mathrm{V}_\beta}$. Therefore, the instantaneous signal-to-interference-plus-noise ratio (SINR) for decoding  ${x_\beta}$ at ${\mathrm{V}_\alpha}$ and BS can be respectively represented as
	\begin{equation}
	r_{\beta\alpha}=\dfrac{P_\beta|h_{\beta\alpha}|^2  d^{-v}_{\beta\alpha}}{P_{SI}|h_{\alpha\alpha}|^2 +\sigma^{2}_n},r_{\beta b}=\dfrac{(1-\lambda)P_\alpha|h_{\alpha b}|^2 d^{-v}_{\alpha b}}{\lambda P_\alpha|h_{\alpha b}|^2 d^{-v}_{\alpha b} +\sigma^{2}_n}.
	\end{equation}

	The instantaneous SINR decoding ${x_\alpha}$ at the BS can be given by 
	\begin{equation}
		r_{\alpha b}=\dfrac{\lambda P_\alpha|h_{\alpha b}|^2 d^{-v}_{\alpha b}}{\sigma^{2}_n}.
	\end{equation} 

	Since the working mode of  ${\mathrm{V}_\alpha}$ is FD, we consider the worst case, that is, ${\mathrm{V}_e}$ can use the maximum ratio combining (MRC) \cite{b43} and NOMA principles to jointly decode the information of ${\mathrm{V}_\beta}$ by simultaneously stealing the transmission information of ${\mathrm{V}_\beta}$ and ${\mathrm{V}_\alpha}$. The instantaneous SINR of ${\mathrm{V}_e}$ decoding ${x_\beta}$ and ${x_\alpha}$ are respectively written by
	\begin{equation}
	r_{\beta e}=\dfrac{P_\beta|h_{\beta e}|^2 d^{-v}_{\beta e}+(1-\lambda)P_\alpha|h_{\alpha e}|^2d^{-v}_{\alpha e}}{\lambda P_\alpha|h_{\alpha e}|^2 d^{-v}_{\alpha e} +P_B d^{-v}_{b e}|\mathbf{w}^H\mathbf{h}_{b e}|^2+\sigma^{2}_n},
	\end{equation}
	\begin{equation}
	r_{\alpha e}=\dfrac{\lambda P_\alpha|h_{\alpha e}|^2 d^{-v}_{\alpha e}}{P_B d^{-v}_{b e}|\mathbf{w}^H\mathbf{h}_{b e}|^2+\sigma^{2}_n}.
    \end{equation}
	
    Based on the definition of secure transmission rate which is the positive discrepancy between the transmission rate and the wiretap rate in \cite{b44}, the secure transmission rate of ${\mathrm{V}_\alpha}$ and ${\mathrm{V}_\beta}$ can be expressed
	as $C_{\beta\alpha}=[R_{\beta\alpha}-R_{\beta e}]^+$ and $C_{\beta b}=[R_{\beta b}-R_{\beta e}]^+$, respectively. According to the principle of decoding in DF, the SINR of decoding information at BS is $ r_\beta=min(r_{\beta\alpha},r_{\beta b})$. 
	
Then the achievable secure transmission rate of ${\mathrm{V}_\beta}$ and ${\mathrm{V}_\alpha}$ are respectively given by
	\begin{equation}
	C_\beta=[min(R_{\beta\alpha},R_{\beta b})-R_{\beta e}]^+, \ \ 	C_\alpha=[R_\alpha-R_{\alpha e}]^+,		
	\end{equation}
 where $ R_q = \mathrm{log}_2(1+r_q) $, $ q \in \left\{\beta\alpha,\beta e, \beta b, \alpha b , \alpha e\right\} $ (in bit/sec/Hz) is the achievable transmission rates of ${\mathrm{V}_o, o \in \left\{\alpha,\beta\right\}}$. 
 
 Then, the offloading and execution time  of  ${\mathrm{V}_o, \  o \in \left\{\alpha,\beta\right\}}$ can be respectively calculated by 
 	\begin{equation}
 	\small
 	D^{off}_o=\frac{(M-m_o)s}{BC_o}, \ D^{exe}_o=\frac{(M-m_o)cs}{f_{MEC}}, \ o \in \left\{\alpha,\beta\right\},	
 	\end{equation}
 where $f_{MEC}$ is the total computing resources allocated by MEC to each group and $ B $ is the bandwidth. Since the return time of the results is very small compared with the offloading and processing time, which can be ignored. So the delay of  ${\mathrm{V}_o, \ o \in \left\{\alpha,\beta\right\}}$ which is required to complete the tasks by MEC is $D^{MEC}_o=D^{off}_o+D^{exe}_o, \ o \in \left\{\alpha,\beta\right\}$.
	
 From the above derivation, it can be known that the time for ${\mathrm{V}_\beta}$ to process the task by MEC is $D^{MEC}_\beta=max(D^{MEC}_{\beta\alpha},D^{MEC}_{\beta b})$. Therefore, the completion time of all tasks on ${\mathrm{V}_\beta}$ is $D_\beta=max(D^{Local}_\beta,D^{MEC}_\beta)$.	

 \section{Secrecy Outage Probability of the System}	
According to \cite{b45}, the SOP is an important indicator to evaluate the security performance of the networks. The SOP of ${\mathrm{V}_\alpha}$ and  ${\mathrm{V}_\beta}$ only reflect their own security, but the security of overall system cannot be completely characterized. In this case, we use the SOPS to evaluate the security of overall networks.  

In this paper, we denote the SOP of ${\mathrm{V}_\alpha}$ and  ${\mathrm{V}_\beta}$ as $P_{sop,\alpha}$ and $\mathit{P}_{sop,\mathit{\beta}}$, respectively. Therefore, the SOPS can be calculated by 
\begin{equation}
	P_{sops}= P_{sop,\alpha}+P_{sop,\beta}-{P_{sop,\alpha}}{ P_{sop,\beta}}.
\end{equation} 
In the following we derive the SOP of ${\mathrm{V}_\alpha}$ and ${\mathrm{V}_\beta}$ sequentially.

\subsection{SOP of ${\mathrm{V}_\alpha}$}
From the Eq.(11), the SOP of ${\mathrm{V}_\alpha}$ is given by
\begin{equation}
	P_{sop,\alpha}=P \left\{C_\alpha<R_s\right\} =P\left\{\frac{1+r_{\alpha b}}{1+r_{\alpha e}}<2^{R_s}\right\},
\end{equation}
where $ R_s $  denoting the target secrecy data rate of vehicles.
Letting $ X=r_{\alpha b}, Y=r_{\alpha e}, Z=(1+X)/(1+Y), X=ZY+Z-1 $, Eq. (13) can be written as
{\setlength\abovedisplayskip{0.1cm}
	\setlength\belowdisplayskip{-0.15cm}
\begin{equation}
	P_{sop,\alpha}=\int_{0}^\infty{F_X(2^{R_s}y+2^{R_s}-1)f_Y(y)\mathrm{d}y}.
\end{equation}}

We can konw that $ |h_{\alpha b}|^2 $ and $ |h_{\alpha e}|^2 $  obeys the exponential distribution with parameter $ \gamma_{\alpha b} $ and $ \gamma_{\alpha e} $ respectively from subsection \RNum{2}-C. Hence, the CDF of \textit{X} in accordance with the Eq. (8) can be obtained as
{\setlength\abovedisplayskip{0.1cm}
\setlength\belowdisplayskip{-0.25cm}
\begin{equation}
	F_X(x)=1-e^{-\frac{\gamma_{\alpha b}x\sigma^{2}_n}{\lambda P_\alpha d^{-v}_{\alpha b}}}.
\end{equation}}

According to Eq. (9), the CDF of \textit{Y} can be calculated by
{\setlength\abovedisplayskip{0.2cm}
	\setlength\belowdisplayskip{-0.2cm}
\begin{equation} 
	\begin{split}
		F_Y(y)&=P\left\{\frac{\lambda P_\alpha|h_{\alpha e}|^2 d^{-v}_{\alpha e}}{P_B d^{-v}_{b e}|\mathbf{w}^H\mathbf{h}_{be}|^2+\sigma^{2}_n}\leq y\right\}\\
		&=P\left\{|h_{\alpha e}|^2\leq \frac{y(P_b d^{-v}_{b e}H_{be}+\sigma^{2}_n)}{\lambda P_\alpha d^{-v}_{\alpha e}}\right\},
	\end{split}
\end{equation} }  

\noindent where $ |\mathbf{w}^H\mathbf{h}_{be}|^2 = \frac{1}{K-1}H_{be}$ can be obtained by Eq. (37) in Section IV. Let $Y_1=|h_{\alpha e}|^2 $ and $Y_2=H_{be}$. 

 In the light of the Eq. (37), $ H_{be} $ is the sum of $ K-1 $ independent random variables following the exponential distribution of parameter $ \gamma_{b e} $. As a result, the PDF of $ Y_2 $ can be calculated by
{\setlength\abovedisplayskip{0.1cm}
	\setlength\belowdisplayskip{-0.35cm}
\begin{equation}
	f_{Y_2}(y_2)=\frac{-\gamma^{K-1}_{b e}}{(K-2)!}y^{K-2}_2 e^{-\gamma_{b e}y_2}.
\end{equation}}

\noindent Then, we can further obtain $ F_Y(y) $ as
{\setlength\abovedisplayskip{0.1cm}
	\setlength\belowdisplayskip{-0.25cm}
\begin{equation}
	\begin{split}
		F_Y(y)&=1-\int_{0}^\infty{e^{-\frac{\gamma_{\alpha e}y(P_b d^{-v}_{b e}y_2+\sigma^{2}_n)}{\lambda P_\alpha d^{-v}_{\alpha e}}}f_{Y_2}(y_2)\mathrm{d}y_2} \\
		&=1-(\frac{-\gamma_{\alpha e}P_b d^{-v}_{b e}y}{\lambda P_\alpha d^{-v}_{\alpha e}\gamma_{b e}}+1)^{1-K}e^{-\frac{\sigma^{2}_n\gamma_{\alpha e}y}{\lambda P_\alpha d^{-v}_{\alpha e}}}. 
	\end{split}
\end{equation}}

\noindent According to the Eq. (18), the PDF of $ Y $ can be obtained as
{\setlength\abovedisplayskip{0.1cm}
	\setlength\belowdisplayskip{-0.25cm}
\begin{equation}
	\begin{split}
		&f_Y(y)=\frac{\sigma^{2}_n\gamma_{\alpha e}}{\lambda P_\alpha d^{-v}_{\alpha e}}\left(\frac{-\gamma_{\alpha e}P_b d^{-v}_{b e}y}{\lambda P_\alpha d^{-v}_{\alpha e}\gamma_{b e}}+1\right)^{1-K}e^{-\frac{\sigma^{2}_n\gamma_{\alpha e}y}{\lambda P_\alpha d^{-v}_{\alpha e}}}\\
		&+(K-1)\frac{\gamma_{\alpha e}P_b\mathit{d}^{-\mathit{v}}_{b e}}{\lambda P_\alpha\mathit{d}^{-\mathit{v}}_{\alpha e}\gamma_{b e}}\left(\frac{-\gamma_{\alpha e}P_b d^{-v}_{b e}y}{\lambda P_\alpha d^{-v}_{\alpha e}\gamma_{b e}}+1\right)^{-K}e^{-\frac{\sigma^{2}_n\gamma_{\alpha e}y}{\lambda P_\alpha\mathit{d}^{-\mathit{v}}_{\alpha e}}}.
	\end{split}
\end{equation}}

\noindent Substituting Eqs. (15) and (19) into Eq. (14), the $ P_{sop,\alpha} $ can be further transformed as
{\setlength\abovedisplayskip{0.1cm}
	\setlength\belowdisplayskip{-0.25cm}
\begin{equation}
	\begin{split}
		P_{sop,\alpha}&=1-e^{-\psi_1}\int_{0}^\infty{e^{-\frac{\sigma^{2}_n\gamma_{\alpha b}zy}{\lambda P_\alpha d^{-v}_{\alpha b}}}f_Y(y)\mathrm{d}y} \\
		&=1-{(K-1)} e^{-\psi_1} \left(\frac{\lambda P_\alpha d^{-v}_{\alpha e}\gamma_{b e}}{P_b d^{-v}_{b e}\gamma_{\alpha e}}\right)^{K-1} \times \\
		&\underbrace{\int_{0}^\infty{e^{-\frac{\sigma^{2}_n\gamma_{\alpha b}zy}{\lambda P_\alpha d^{-v}_{\alpha b}}-\frac{\sigma^{2}_n\gamma_{\alpha e}y}{\lambda P_\alpha d^{-v}_{\alpha e}} }\left(y+\frac{\lambda P_\alpha d^{-v}_{\alpha e}\gamma_{b e}}{P_b d^{-v}_{b e}\gamma_{\alpha e}}\right)^{-K}}\mathrm{d}y}_{E_1}-\\
		&\frac{\sigma^{2}_n\gamma_{\alpha e}e^{-\psi_1}}{\lambda P_\alpha d^{-v}_{\alpha e}}\left(\frac{\lambda P_\alpha d^{-v}_{\alpha e}\gamma_{b e}}{P_b d^{-v}_{b e}\gamma_{\alpha e}}\right)^{K-1}\times \\
		&\underbrace{\int_{0}^\infty{e^{-\frac{\sigma^{2}_n\gamma_{\alpha b}zy}{\lambda P_\alpha d^{-v}_{\alpha b}}-\frac{\sigma^{2}_n\gamma_{\alpha e}y}{\lambda P_\alpha d^{-v}_{\alpha e}} }\left(y+\frac{\lambda P_\alpha d^{-v}_{\alpha e}\gamma_{b e}}{P_b d^{-v}_{b e}\gamma_{\alpha e}}\right)^{1-K}}\mathrm{d}y}_{E_2},
	\end{split}	
\end{equation}}

\noindent where $ \psi_1=\frac{\sigma^{2}_n\gamma_{\alpha b}(2^{R_s}-1)}{\lambda P_\alpha d^{-v}_{\alpha b}} $. Then, $ E_1 $ can be calculated by
{\setlength\abovedisplayskip{0.1cm}
	\setlength\belowdisplayskip{-0.25cm}
\begin{small}
	\begin{equation}
		\begin{split}
			E_1&=\frac{1}{(K-1)!}\sum_{i=1}^{K-1}(i-1)!\left(-\frac{\sigma^{2}_n\gamma_{\alpha b}z}{\lambda P_\alpha d^{-v}_{\alpha b}}-\frac{\sigma^{2}_n\gamma_{\alpha e}}{\lambda P_\alpha d^{-v}_{\alpha e}}\right)^{K-i-1}\times \\
			&{\left(\frac{\lambda P_\alpha d^{-v}_{\alpha e}\gamma_{b e}}{P_b d^{-v}_{b e}\gamma_{\alpha e}}\right)}^{-i} 
			-\frac{1}{(K-1)!}\left(-\frac{\sigma^{2}_n\gamma_{\alpha b}z}{\lambda P_\alpha d^{-v}_{\alpha b}}-\frac{\sigma^{2}_n\gamma_{b e}}{\lambda P_\alpha d^{-v}_{\alpha e}}\right)^{K-1}\times \\
			&e^{\left(-\frac{\sigma^{2}_n\gamma_{\alpha b}z}{\lambda P_\alpha d^{-v}_{\alpha b}}-\frac{\sigma^{2}_n\gamma_{\alpha e}}{\lambda P_\alpha d^{-v}_{\alpha e}}\right)\frac{\lambda P_\alpha d^{-v}_{\alpha e}\gamma_{b e}}{P_b d^{-v}_{b e}\gamma_{\alpha e}}}\times \\
			&E_i\left[\left(\frac{\lambda P_\alpha d^{-v}_{\alpha e}\gamma_{\alpha e}}{P_b d^{-v}_{b e}\gamma_{\alpha e}}-\frac{\sigma^{2}_n\gamma_{\alpha b}z}{\lambda P_\alpha d^{-v}_{\alpha b}}\right)\frac{\lambda P_\alpha d^{-v}_{\alpha e}\gamma_{b e}}{P_b d^{-v}_{b e}\gamma_{\alpha e}}\right] \\
			=&\frac{1}{(K-1)!}\left(\frac{\lambda P_\alpha}{\sigma^{2}_n}\right)^{1-K}\sum_{i=1}^{K-1}(i-1)!(-\psi_2)^{K-i-1}\psi_3^{-i}-\\
			&\frac{1}{(K-1)!}\left(\frac{\lambda P_\alpha}{\sigma^{2}_n}\right)^{1-K}\psi_2^{K-1}e^{-\psi_2\psi_3}E_i(-\psi_2\psi_3),
		\end{split}	
	\end{equation} 
\end{small}}

\noindent where $ \psi_2=-\frac{\gamma_{\alpha e}}{\mathit{d}^{-\mathit{v}}_{\alpha e}}-\frac{\gamma_{\alpha b}z}{\mathit{d}^{-\mathit{v}}_{\alpha b}} $, $ \psi_3=\frac{\sigma^{2}_n\mathit{d}^{-\mathit{v}}_{\alpha e}\gamma_{b e}}{\mathit{P}_b\mathit{d}^{-\mathit{v}}_{b e}\gamma_{\alpha e}} $.
Similar to solving $ E_1 $, we can calculate $ E_2 $ as
\begin{small}
	\begin{equation}
		\begin{split}
			E_2=&\frac{1}{(K-2)!}\left(\frac{\lambda P_\alpha}{\sigma^{2}_n}\right)^{2-K}\sum_{i=1}^{K-2}(i-1)!(-\psi_2)^{K-i-2}\psi_3^{-i}-\\
			&\frac{1}{(K-2)!}\left(\frac{\lambda P_\alpha}{\sigma^{2}_n}\right)^{2-K}\psi_2^{K-2}e^{-\psi_2\psi_3}E_i(-\psi_2\psi_3).
		\end{split}	
	\end{equation} 
\end{small}

Substituting Eqs. (21), (22) into Eq. (20), $ P_{sop,\alpha} $ be acquired by 
\begin{equation}
\small
	\begin{split}
		&P_{sop,\alpha}=\int_{0}^\infty{\left[	1-e^{-\frac{\gamma_{\alpha b}(2^{R_s}y+2^{R_s}-1)}{\lambda P_\alpha d^{-v}_{\alpha b}}}\right]f_Y(y)\mathrm{d}y}\\	
		&=1-e^{-\psi_3}+\frac{e^{-\psi_3}2^{R_s}}{(K-2)!}\frac{\gamma_{\alpha b}}{d^{-v}_{\alpha b}}  \left( \frac{P_b\mathit{d}^{-v}_{b e}\gamma_{\alpha e}}{\sigma^{2}_n d^{-v}_{\alpha e}\gamma_{b e}}\right)^{1-K} \times \\
		&\left[\sum_{i=1}^{K-2} (i-1)!{\psi_1}^{K-i-2}{\psi_2}^{-i}-{\psi_1}^{K-2}e^{-\psi_1\psi_2}E_i(\psi_1\psi_2)\right].
	\end{split}	
\end{equation} 

\newcounter{TempEqCnt2}
\setcounter{TempEqCnt2}{\value{equation}} 
\setcounter{equation}{30} 
\begin{figure*}[hb] 
	\hrulefill 
	\begin{equation}
	\begin{split}
	F_Y(y)=&\int_{0}^\infty\int_{0}^{\frac{yP_bd^{-v}_{be}y_3+y\sigma^2_n}{(y\lambda-1+\lambda)P_\alpha d^{-v}_{\alpha e}}}{F_{Y_1}(\theta_1)f_{Y_2}(y_2)f_{Y_3}(y_3)\mathrm{d}y_2\mathrm{d}y_3} \\
	=&1-\frac{\gamma^{K-1}_{be}}{(K-2)!}e^{-\frac{y\sigma^2_n\gamma_{\alpha e}}{(1-\lambda-y\lambda)P_\alpha d^{-v}_\alpha e}}\int_{0}^\infty{e^{-\frac{yP_bd^{-v}_{be}y_3}{(1-\lambda-y\lambda)P_\alpha d^{-v}_{\alpha e}}-\gamma_{b e}y_3}y^{K-2}_3\mathrm{d}y_3}-\\
	&\frac{\gamma^{K-1}_{be}}{(K-2)!}\left(\frac{(y\lambda-1+\lambda)P_\alpha d^{-v}_{\alpha e}\gamma_{\beta e}}{P_\beta d^{-v}_{\beta e}\gamma_{\alpha e}}+1\right)^{-1} e^{-\frac{\gamma_{\beta e}\sigma^2_n y}{P_\beta d^{-v}_{\beta e}}}\int_{0}^\infty{e^{-\frac{P_b d^{-v}_{be}\gamma_{\beta e}y y_3}{P_\beta d^{-v}_{\beta e}}-\gamma_{b e}y_3}y^{K-2}_3\mathrm{d}y_3}-\\
	&\frac{\gamma^{K-1}_{be}}{(K-2)!}\left(\dfrac{(y\lambda-1+\lambda)P_\alpha d^{-v}_{\alpha e}\gamma_{\beta e}}{P_\beta d^{-v}_{\beta e}\gamma_{\alpha e}}+1\right)^{-1}e^{-\frac{y\sigma^2_n\gamma_{\alpha e}}{(1-\lambda-y\lambda)P_\alpha d^{-v}_\alpha e}}\int_{0}^\infty{e^{-\dfrac{P_b d^{-v}_{be}\gamma_{\alpha e}y y_3}{(1-\lambda-y\lambda)P_\alpha d^{-v}_\alpha e}-\gamma_{b e}y_3}y^{K-2}_3\mathrm{d}y_3} \\
	=&1-e^{-\theta_4 y} \left[1-\frac{ \theta_4\lambda(\tau_1-y)}{\theta_2}\right]^{-1}  \left(1+\frac{\theta_4}{\theta_3} y \right)^{1-K}
	- e^{-\frac{\theta_2 y }{ \lambda(\tau_1-y)}} \left[1-\frac{\theta_2}{ \theta_4\lambda(\tau_1-y)}\right]^{-1} \left[1+\frac{\theta_2 y }{ \theta_3\lambda(\tau_1-y)}\right]^{1-K},
	\end{split}	
	\end{equation}
	where $ \theta_1=\frac{(y\lambda-1+\lambda)P_\alpha d^{-v}_{\alpha e}y_2+yP_bd^{-v}_{be}y_3+y\sigma^2_n}{P_\beta d^{-v}_{\beta e}} $, $ \theta_2= \frac{\sigma^2_n \gamma_{\alpha e}}{P_\alpha d^{-v}_{\alpha e}}$, $ \theta_3= \frac{\sigma^2_n \gamma_{b e}}{P_b d^{-v}_{b e}} $, $ \theta_4= \frac{ \sigma^2_n \gamma_{\beta e}}{P_\beta d^{-v}_{\beta e}} $. 
\end{figure*}
\setcounter{equation}{\value{TempEqCnt2}}

\subsection{SOP of ${\mathrm{V}_\beta}$}
From the Eq. (10), the SOP of ${\mathrm{V}_\beta}$ can be obtained as
\begin{equation}
	\begin{split}
		\mathit{P}_{sop,\beta}=P\left\{\frac{1+min(r_{\beta \alpha},r_{\beta b})}{1+r_{\beta e}}<2^{R_s}\right\}.
	\end{split}
\end{equation}
Letting $ X=min(r_{\beta \alpha},r_{\beta b}), Y=r_{\beta e}, Z=(1+X)/(1+Y), X=ZY+Z-1 $, Eq. (24) can be written as the Eq. (14).
In the view of theory of probability, the CDF of \textit{X} can be obtained as
\begin{equation}
	\begin{split}
		F_X( x )&=P\left\{X\leq x\right\}=1-P\left\{X \geq x\right\} \\
		&=1-P\left\{min(r_{\beta \alpha},r_{\beta b})\geq x\right\}\\
		&=1-[1-F_{r_{\beta \alpha}}(x)][1-F_{r_{\beta b}}(x)].
	\end{split}
\end{equation}

According to the Eq. (6), the CDF of $ r_{\beta \alpha} $ can be written as
\begin{equation}
	\begin{split}
		F_{r_{\beta \alpha}}(x)& =P\left\{\frac{P_\beta |h_{\beta\alpha}|^2  d^{-v}_{\beta\alpha}}{P_{SI}|h_{\alpha\alpha}|^2 +\sigma^{2}_n} \leq x\right\} \\
		&=P\left\{|h_{\beta\alpha}|^2 \leq \frac{P_{SI}|h_{\alpha\alpha}|^2 + \sigma^{2}_n}{P_\beta d^{-v}_{\beta\alpha}} x\right\}.
	\end{split}
\end{equation}
Similar to the Eq. (16), $ F_{r_{\beta \alpha}}(x) $ can be calculated by
\begin{equation}
\small
	\begin{split}	
		F_{r_{\beta \alpha}}(x)& =1- \int_{0}^\infty{e^{-\gamma_{\beta\alpha}x\left(\frac{P_{SI}x_2 + \sigma^{2}_n}{P_\beta d^{-v}_{\beta\alpha}}\right)}\gamma_{\alpha\alpha}e^{-\gamma_{\alpha\alpha}x_2}\mathrm{d}x_2} \\
		&=1-\frac{P_\beta d^{-v}_{\beta\alpha}\gamma_{\alpha\alpha}}{P_{SI}\gamma_{\beta\alpha}+ P_\beta d^{-v}_{\beta\alpha}\gamma_{\alpha\alpha}}e^{-\frac{\gamma_{\beta\alpha}\sigma^{2}_n x}{ P_\beta d^{-v}_{\beta\alpha}}}.
	\end{split}
\end{equation}
On the basis of the Eq. (6), $F_{r_{\beta b}}(x) $ can be acquired by
\begin{equation}
	\begin{split}
		F_{r_{\beta b}}(x)&=P\left\{\frac{(1-\lambda)P_\alpha|h_{\alpha b}|^2 d^{-v}_{\alpha b}}{\lambda P_\alpha|h_{\alpha b}|^2 d^{-v}_{\alpha b} +\sigma^{2}_n} \leq x\right\}\\
		&=P\left\{(1-\lambda-x\lambda)P_\alpha d^{-v}_{\alpha b}|h_{\alpha b}|^2 \leq x \sigma^{2}_n\right\}\\
		&=
		\begin{cases}
			1-e^{-\frac{\gamma_{\alpha b}\sigma^{2}_n x}{\lambda\mathit{P}_\alpha \mathit{d}^{-\mathit{v}}_{\alpha b}}(	\tau_1-x)}, & x\leq \tau_1 \\
			1\ \ \ \ \ \ \ \ \ \ \ \ \ \ \ \ \ \ \ \ \ \ ,& x\geq\tau_1  \\
		\end{cases},
	\end{split}
\end{equation}
where $ \tau_1=\frac{1-\lambda}{\lambda} $. Substituting Eqs. (27), (28) into Eq. (25), $ F_X(x) $ can be further expressed as
\begin{equation}
	F_X(x)=  \begin{cases}
		1-\phi(x), & x \leq\tau_1 \\
		1\ \ \ \ \ \ \ \ \ ,& x\geq\tau_1  \\
	\end{cases},
\end{equation}
where $ \phi(x)=\frac{P_\beta d^{-v}_{\beta\alpha}\gamma_{\alpha\alpha}}{P_{SI}\gamma_{\beta\alpha}+ P_\beta d^{-v}_{\beta\alpha}\gamma_{\alpha\alpha}}e^{-\frac{\gamma_{\beta\alpha}\sigma^{2}_n x}{ P_\beta d^{-v}_{\beta\alpha}} }$.

In accordance with the Eq. (8), $ F_Y(y) $ can be calculated by
\begin{equation}
    \small
	\begin{split}
		&F_Y(y)=P\left\{\frac{P_\beta|\mathit{h}_{\beta e}|^2 d^{-v}_{\beta e}+(1-\lambda)P_\alpha|h_{\alpha e}|^2 d^{-v}_{\alpha e}}{\lambda P_\alpha|h_{\alpha e}|^2 d^{-v}_{\alpha e} +{P_B d^{-v}_{b e}|\mathbf{w}^H\mathbf{h}_{b e}|^2+\sigma^{2}_n}}\leq y\right\}\\ 
		&=P\left\{|h_{\beta e}|^2\leq\frac{(y\lambda-1+\lambda)P_\alpha d^{-v}_{\alpha e}|h_{\alpha e}|^2+y(P_bd^{-v}_{be}H_{be}+\sigma^2_n)}{P_\beta d^{-v}_{\beta e}}\right\}.
	\end{split}
\end{equation}
Let $ Y_1=|\mathit{h}_{\beta e}|^2 $, $ Y_2=|h_{\alpha e}|^2 $, $ Y_3=H_{be} $. 
Then, the CDF of \textit{Y} can be acquired by (31), which is displayed at the bottom of this page. Then $f_Y(y)$ can be derived from $  F_Y(y) $. 

On account of $ F_X(x) $ and $ f_Y(y) $, we can further transform the $ P_{sop,\beta} $ as
\setcounter{equation}{31}
\begin{equation}
	\begin{split}
		&P_{sop,\beta}=1-\int_{0}^{\tau_2}{\phi(2^{R_s}y+2^{R_s}-1)f_Y(y)\mathrm{d}y}\\
		&=1-\frac{\tau_2}{2}\int_{-1}^{1}{\phi[z\dfrac{\tau_2}{2}(t+1)+z-1]f_Y[\frac{\tau_2}{2}(t+1)]\mathrm{d}t},
	\end{split}
\end{equation}
where $ y=\frac{\tau_2}{2}(t+1) $ and $ \tau_2 = \frac{1-z\lambda}{z\lambda} $. However, obtaining the closed-form expressions of $ \mathit{P}_{sop,\beta} $ is difficult. As a resort to reduce the computational complexity, Gaussian-Legendre quadrature \cite{b46} can be employed in this paper.
By Gaussian-Laguerre quadrature, $ P_{sop,\beta} $ can be approximated by
\begin{equation}
	\begin{split}	
	P_{sop,\beta}&\approx 1-\frac{\tau_2}{2}\sum_{i=1}^{N_\beta} A_i f_Y\left[\frac{\tau_2}{2}(t_i+1)\right] \\
		&\times \phi\left[2^{R_s}\frac{\tau_2}{2}(t_i+1)+2^{R_s}-1\right], 
	\end{split}
\end{equation}
where $ A_i=\frac{2}{(1-t^2_i)[P^{\prime}_{n}(t_i)]^2} $ and $ P_n(t)=\frac{1}{2^n(n)!}\frac{d^n}{dtn}(t^2-1)^n$, $ t_i $ is the \textit{i}-th root of $ P_{n+1}(t) $.   

Till now, substituting Eqs. (23) and (33) into Eq. (12), we can get an approximate closed form of SOPS by
\begin{equation}
\small
	\begin{split}
&P_{sops}\approx 2-e^{-\psi_3}+\frac{e^{-\psi_3}2^{R_s}}{(K-2)!}\frac{\gamma_{\alpha b}}{d^{-v}_{\alpha b}}  \left( \frac{P_b\mathit{d}^{-v}_{b e}\gamma_{\alpha e}}{\sigma^{2}_n d^{-v}_{\alpha e}\gamma_{b e}}\right)^{1-K} \times \\
&\left[\sum_{i=1}^{K-2} (i-1)!{\psi_1}^{K-i-2}{\psi_2}^{-i}-{\psi_1}^{K-2}e^{-\psi_1\psi_2}E_i(\psi_1\psi_2)\right]-\\
&\frac{\tau_2}{2}\sum_{i=1}^nA_i\phi\left[2^{R_s}\dfrac{\tau_2}{2}(t_i+1)+2^{R_s}-1\right]\times f_Y\left[\frac{\tau_2}{2}(t_i+1)\right]- \\
&\left\{1-e^{-\psi_3}+\frac{e^{-\psi_3}2^{R_s}}{(K-2)!}\frac{\gamma_{\alpha b}}{d^{-v}_{\alpha b}}  \left( \frac{P_b\mathit{d}^{-v}_{b e}\gamma_{\alpha e}}{\sigma^{2}_n d^{-v}_{\alpha e}\gamma_{b e}}\right)^{1-K} \times \right.\\
&\left. \left[\sum_{i=1}^{K-2} (i-1)!{\psi_1}^{K-i-2}{\psi_2}^{-i}-{\psi_1}^{K-2}e^{-\psi_1\psi_2}E_i(\psi_1\psi_2)\right]\right\}\times\\
&\left\{1-\frac{\tau_2}{2}\sum_{i=1}^nA_i\phi\left[2^{R_s}\frac{\tau_2}{2}(t_i+1)+2^{R_s}-1\right]  f_Y\left[\dfrac{\tau_2}{2}(t_i+1)\right]\right\}. 
	\end{split}
\end{equation} 
	
    \section{Problem Formulation}
    In this section, the optimization problem of minimizing $D_\beta$ is tackled by jointly optimizing the vehicle task division, power allocation and the transmit beamforming, which is mathematically formulated as
    \setlength{\textfloatsep}{20pt}
    \begin{equation}
    \small
    \begin{aligned} 
    \label{P1}
    &\mathcal{P}_1:\min\limits_{\mathbf{w},\lambda,m_\alpha,m_\beta}\quad D_\beta=max \left\{D^{Local}_\beta,D^{MEC}_{\beta\alpha},D^{MEC}_{\beta b}\right\}  \vspace{0.7ex} \\
    &\begin{array}{r@{\quad}r@{}l@{\quad}l}
    s.t. &\mathrm{C}1: & \lambda \in (0,0.5),  \vspace{0.7ex}\\
    &\mathrm{C}2: & m_o\in\left\{0,1,2,\ldots,M\right\}, \   o=\left\{\alpha,\beta\right\}, \vspace{0.7ex}\\
    &\mathrm{C}3: & 0\leq D^{Local}_o=\frac{m_ocs}{f_{Local}}\leq D,  \  o=\left\{\alpha,\beta\right\}, \vspace{0.7ex}\\
    &\mathrm{C}4: &0\leq D^{MEC}_\alpha=\frac{(M-m_\alpha)s}{BC_\alpha}+\frac{(M-m_\alpha)cs}{f_{MEC}}\leq D,  \vspace{0.7ex}\\
    &\mathrm{C}5: & 0\leq D^{MEC}_{\beta\alpha}=\frac{(M-m_\beta)s}{BC_{\beta\alpha}}+\frac{(M-m_\beta)cs}{f_{MEC}}\leq D, \vspace{0.7ex}\\
    &\mathrm{C}6: & 0\leq D^{MEC}_{\beta b}=\frac{(M-m_\beta)s}{BC_{\beta b}}+\frac{(M-m_\beta)cs}{f_{MEC}}\leq D,  \vspace{0.7ex}\\
    &\mathrm{C}7: & P_{sop,o}\leq \zeta,  \  o=\left\{\alpha,\beta\right\}, \vspace{0.7ex}\\
    &\mathrm{C}8: & \mathbf{w}^H\mathbf{h}=\mathbf{0}, \vspace{0.7ex}\\
    &\mathrm{C}9: & \mathbf{w}^H\mathbf{w}\leq 1. \\         
    \end{array}
    \end{aligned}
    \end{equation}
     Constraint in (C1) specifies the range of the PAR, while constraint in (C2) describes the number of the task executed locally. Moreover, constraints in (C3--C6) guarantee that the total task completion delay of each vehicle is limited to $D$. Constraint in (C7) shows the maximum tolerable SOP $ \zeta $ for each legal vehicle. The constraint in (C8) and (C9) have been introduced in subsection II--C. 
    
    It can be seen from $ \mathcal{P}_1 $ that the optimization problem involves three one-dimensional variables and one $ (K-1)\times 1$ dimensional vector variable. Considering that the influence of $ \mathbf{w} $ on the system is relatively independent, in order to simplify the analysis, we first consider the optimization of $ \mathbf{w} $. When only considering the impact of eavesdropping rate, $ \mathbf{w} $  is related to the strength of AN's interference to $ V_e $. The smallest time delay for task completion of $ V_\beta $ is the smallest eavesdropping rate, that is, the largest interference, $ \mathcal{P}_1 $ can be rewritten as

    \begin{equation}
    	\begin{aligned} \label{P2}
    		&\mathcal{P}_2:\max\limits_{\mathbf{w}}\quad |\mathbf{w}^H\mathbf{h}_{be}|^2 \vspace{1.5ex}\\
    		&\begin{array}{r@{\quad}r@{}l@{\quad}l}
    			s.t. \quad\mathrm{C}8,\mathrm{C}9.       
    		\end{array}
    	\end{aligned}
    \end{equation}
    
    It is easy to know from $ \mathcal{P}_2 $ that when $  \mathbf{w}^H\mathbf{w}=1 $, it is possible to get the maximum value of $ |\mathbf{w}^H\mathbf{h}_{be}|^2  $. Let $  \Lambda $ be the projection matrix of the column of $ \mathbf{h} $ on the zero space, then $ \Lambda = \mathbf{I}_{K-1} -\mathbf{h}(\mathbf{h}^H\mathbf{h})^{-1} \mathbf{h}^H$, where $ \mathbf{I}_{K-1} $ is the $ (K-1)\times (K-1)$ dentity matrix. Then any results satisfying C7 also satisfy $ \Lambda \mathbf{w}=\mathbf{w}  $, that is $ \mathbf{w}^H\mathbf{h}_{be}=\mathbf{w}^H\Lambda\mathbf{h}_{be} $, the optimal solution is $ \mathbf{w}^*=\dfrac{\Lambda\mathbf{h}_{be}}{ \left\|\Lambda\mathbf{h}_{be}\right\|} $. Therefore, $ \mathbf{w}^* $ is actually the orthogonal projection of $ \mathbf{h}_{be} $  on the null space of $ \mathbf{H} $. So the optimal solution of $ \mathcal{P}_2 $  can be obtained as
    \begin{equation}
    	|\mathbf{w}^H\mathbf{h}_{be}|^2=\dfrac{1}{K-1} \sum_{k=1}^{K-1}|h_{be,k}|^2=\dfrac{1}{K-1} H_{be},
    \end{equation}
    and the transmission power of a single antenna is $ P_b=\frac{P_B}{K-1} $.
    
    After obtaining $ \mathbf{w}^* $, by introducing an auxiliary variable Z, $ \mathcal{P}_1 $ can be rewritten as	
    \begin{equation}
    	\begin{aligned} \label{P3}
    		&\mathcal{P}_3:\min\limits_{\lambda,m_\alpha,m_\beta}\quad Z=D_\beta \vspace{1.5ex}\\
    		&\begin{array}{r@{\quad}r@{}l@{\quad}l}
    			s.t.
    			&\mathrm{C}1 \sim & \mathrm{C}7, \vspace{1.5ex}\\ 
    			&\mathrm{C}10: &\quad Z \geq D^{local}_\beta, \vspace{1.5ex}\\
    			&\mathrm{C}11: &\quad Z \geq D^{MEC}_j, \ j=\left\{\beta\alpha,\beta b\right\}, \vspace{1.5ex}\\
    			&\mathrm{C}12: &\quad 0\leq Z\leq D. \\       
    		\end{array}
    	\end{aligned}
    \end{equation}
    
    Because the optimization problem  $ \mathcal{P}_3 $ is a MINLP problem, the linear method cannot be used to solve it. Furthermore, in terms of complexity, as the scale of network grows, the MINLP problem is hard to be solved by the exhaustive search approach within an acceptable time. Therefore, we propose the GA-PATS to solve $ \mathcal{P}_3 $. The key idea is survival of the fittest and ultimately selection the best. As a meta-heuristic algorithm, GA is based on Darwinism and Mendel's genetic theory \cite{b47}, which is an efficient and global optimization method that has been widely used in wireless communication network for combinatorial optimization MINLP problem \cite{b48,b49}. The details for GA-PATS algorithm are as follows.   
        
    \IncMargin{1em}
    \begin{algorithm}
    	\setlength{\intextsep}{5pt}
  	\LinesNumbered 
  \caption{GA-Based Power Allocation and Task Scheduling (GA-PATS)} 
  \label{GA-Based Power Allocation and Task Scheduling (GA-PATS)}
  \KwIn{ $  M $, $ P_{\alpha} $, $ P_{\beta} $, $ v $, $ \zeta $, $ D $, $ s $, $ c $, $ f_{Local} $, $ f_{MEC} $, number of populations \textit{S}, number of iterations \textit{I}, the minimum tolerance $ \digamma $ }
  \KwOut{$ \lambda, m_\beta, m_\alpha, D_\beta, D_\alpha $}
  Randomly initialize the population based on (C1), (C2), (C12).\\
  \For{$i= 1:I$}{
  	Calculate $ D^{Local}_\beta,D^{MEC}_{\beta\alpha},D^{MEC}_{\beta b} $ by using (C4), (C5) and (C6), respectively. \\
  	Calculate $ P_{sop,\alpha}$ and $P_{sop,\beta}$ by using Eqs. (23) and (33), respectively. \\
  	\If{The optimization criteria are met}{
  		$ D_\beta(i)=Z$; \\
  	}
  	Compute the penalty fitness value and select individuals by using roulette wheel selection.\\
  	Cross two adjacent individuals with the crossover probability $ \varepsilon $.\\
  	Change individuals with mutation probability $ \iota $. \\
  	Update population.\\		
  	\If{$ i \geq 2 \ \&\& \  |D_\beta(i)-D_\beta(i-1)| \leq \digamma $}{
  		break; \\
  	}
  }
    \end{algorithm}
    \DecMargin{1em}

\section{Simulation Results}	
In this section, numerical results are employed to validate the performance of the system. “RPM” represents the random pairing method that randomly selects a vehicle from group C to pair with the vehicle from group E, while “GPM” is proposed in subsection II--A. Moreover, “EG” represents the exhaustive algorithm, which traverses all possible results to find the optimum, and “GA-PATS” is introduced in section IV. “AN” and “N-AN”  respectively indicate that BS sends and does not send the AN signal.  “Ana.” and “Sim.” represent the analysis and simulation results, respectively. 

The main system simulation parameters are listed in Table I. By default, in the following simulation process, we consider all edge vehicles in the system, ${\mathrm{V}_\beta}$ and ${\mathrm{V}_\alpha}$ represent the edge vehicles and the center vehicles respectively. The simulation results are obtained by averaging over 10000 random channel realizations.
 \begin{table}
	\centering
	\caption{\\
		\bf SIMULATION PARAMETERS}
     \label{tab1}
     \centering
   \begin{tabular}{|l|l|l|}
	\hline 
	\bf Notation & \bf Description  & \bf Value\\
	\hline  
	$ N $ & The number of vehicles &40\\
	\hline 
	$ M $& The number of tasks per vehicle &10\\
    \hline
	$ N_\beta $ & The number of interpolation &  500\\
	\hline 
	$ R $& The radius of BS coverage &500m\\
	\hline 
	$ R_{MC} $ & The radius of the center group range &300m\\
	\hline
	$ K $ & The number of antennas in BS &10\\
	\hline
	$ B $ & Available bandwidth &1MHz\\
	\hline
	$ P_B $ & The maximum power of BS to transmit AN & 40 dBm\\
	\hline
	$ P_\alpha $ & The power of center vehicles & 10 dBm\\
	\hline
	$ N_0 $ & The AWGN spectral density & -174 dBm/Hz\\
	\hline
	$ f_{MEC} $ & The CPU frequency of MEC &  $ 5 \times 10^{10}$  Hz\\
	\hline
	$ f_{Local} $ & The CPU frequency of vehicle & $ 5 \times 10^8$ Hz\\
    \hline
	$ c $ & The number of CPU cycles required per bit &  1000\\
	\hline
	$ s $ & The size of the data contained each task & $ 1 \times 10^5$ bit\\
	\hline
    $ v $ & The path-loss exponent & 3\\ 
    \hline
\end{tabular}
\end{table}

	\begin{figure}[t]
		\setlength{\textfloatsep}{2pt}
	\centering
	\includegraphics[width=3.5in,height=2.5in]{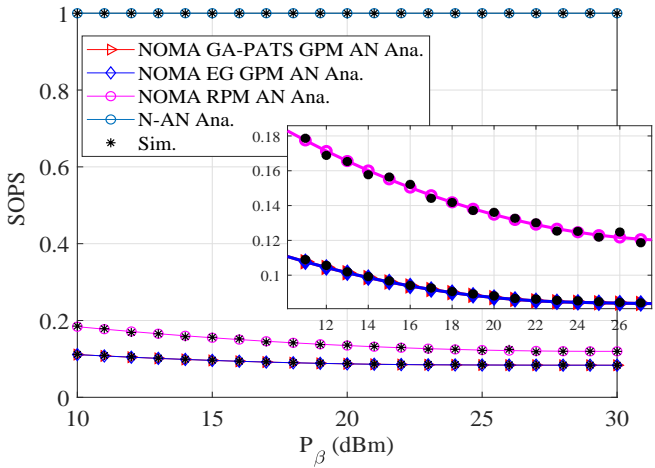} 	
	\caption{The SOPS in different schemes versus $ P_\beta $.}
 	\label{fig3}
   \end{figure}
   Figure \ref{fig3} depicts the SOPS in different schemes versus $ P_\beta $. From Fig. \ref{fig3}, it is observed that the two curves obtained by GA--PATS and EG basically coincide, indicating that the proposed GA-PATS achieves the the optimum exactly. When the $ P_\beta $ is small, with the increase of the power, the SOPS decreases obviously. When the power increases to a threshold, the curve in the figure tends to be stable. This is because when the power is small, the power increase can significantly improve the communication channel conditions. However, the influence of power increase on the communication channel is insignificant under an excellent communication environment. 
   
   In addition, as shown in Fig. \ref{fig3}, the SOPS obtained by RPM is larger than that obtained by GPM, especially in the case of low power consumption, which can prove that our proposed GPM effectively improves the security performance by properly pairing the vehicles near to MEC with the edge vehicles. Besides, the simulation results coincide with the analytical ones, verifying the correctness of the theoretical derivations. Moreover, compared with the results of N-AN scheme, sending AN can significantly improve the security performance of the system. Since the security performance of the schemes without AN is extremely low, these schemes will not be compared below.
   
\setlength{\textfloatsep}{5pt}
	\begin{figure}[t]
	\centering
	\includegraphics[width=3.5in,height=2.5in]{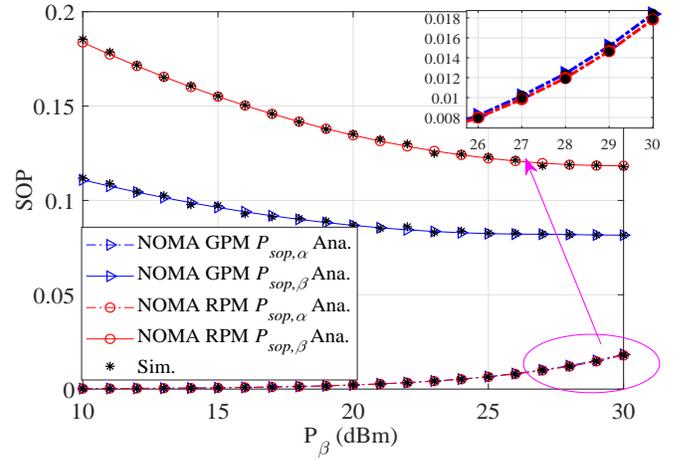}		
	\caption{The SOPs of ${\mathrm{V}_\beta}$ and ${\mathrm{V}_\alpha}$ in diverse schemes versus $ P_\beta $.}
	\label{fig4}
\end{figure}

   Figure \ref{fig4} draws the SOPs of ${\mathrm{V}_\beta}$ and ${\mathrm{V}_\alpha}$ in diverse schemes versus $ P_\beta $. It can be seen that although the $ P_{sop,\alpha}$ of the GPM scenario is larger than the RPM scenario, the gap between them is very close. On the other hand, the $ P_{sop,\beta}$ of the GPM scheme is significantly smaller than that of the RPM scheme. It can be concluded that compared with the RPM scheme, the GPM scheme we proposed achieves much higher safety performance improvement of the edge vehicles at the cost of insignificant loss of safety performance of the center vehicles.
   
   \setlength{\textfloatsep}{5pt}
   \begin{figure}[t]
   	\centering
   	\includegraphics[width=3.5in,height=2.5in]{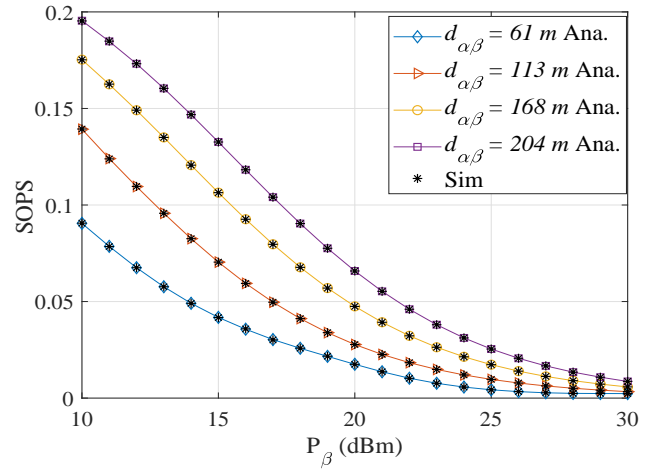}	
   	\caption{The SOPS in diverse straight-line distance between ${\mathrm{V}_\beta}$ and ${\mathrm{V}_\alpha}$ versus $ P_\beta $.}
   	\label{fig5}
   \end{figure} 

   Furthermore, Fig. \ref{fig5} reveals the SOPS of multiple vehicle pairs in the vehicle cluster versus $ P_\beta $ under various straight-line distances between${\mathrm{V}_\beta}$ and ${\mathrm{V}_\alpha}$. As the $ d_{\alpha\beta} $ increases, the value of SOPS also increased due to the worse communication channel condition. In addition, with the increase of $ P_\beta $, the declining trend of SOPS of each scheme gradually slows down, which is consistent with the variation tendency of these schemes using AN to interfere with the potential eavesdropping of vehicles in Fig. \ref{fig3}, and the gap between each curve also gradually decreases in Fig. \ref{fig5}.
 
   \setlength{\textfloatsep}{5pt}
	\begin{figure}[t]
	\centering
	\includegraphics[width=3.5in,height=2.5in]{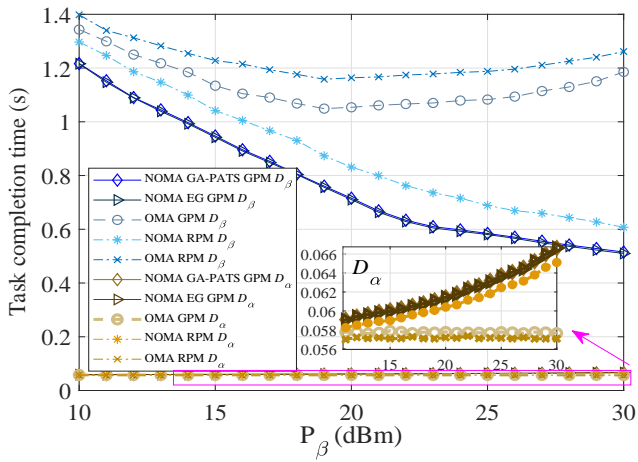}		
	\caption{The task completion time of ${\mathrm{V}_\beta}$ and ${\mathrm{V}_\alpha}$, i.e., $ D_\beta $ and $ D_\alpha $, in different schemes versus $ P_\beta $.}
	\label{fig6}
   \end{figure} 
   
   Figure \ref{fig6} displays the task completion time of ${\mathrm{V}_\beta}$ and ${\mathrm{V}_\alpha}$, i.e., $ D_\beta $ and $ D_\alpha $, under different schemes. Let us check the results for ${\mathrm{V}_\beta}$ first. It is clear that NOMA can effectively reduce task latency of ${\mathrm{V}_\beta}$ compared to OMA. On the other hand, compared with RPM, our proposed GPM is superior in terms of latency performance. Furthermore, Fig. \ref{fig6} also shows that under the NOMA scheme, the task delay of ${\mathrm{V}_\beta}$ decreases with the increase of $P_\beta $. By contrast, an increase in $P_\beta $ after 20 dBm causes a gradual rise in the task delay under the OMA scheme.

   Further analysis reveals that this phenomenon is due to the fact that the power allocation ratio $ \lambda $ is constant in OMA. From Section II,  it can be known that the task completion delay of ${\mathrm{V}_\beta}$ depends on the minimum achievable rate of information transmission in the channel, which is $ R_{\beta} = min(R_{\beta\alpha},R_{\beta b}) $.  The augment of $ P_\beta $ leads to the increase of $ R_{\beta\alpha} $ but diminution of $ R_{\beta b} $. When the $P_\beta $ is small, the channel condition between $ \mathrm{V}_\beta $ and $ \mathrm{V}_\alpha $ is worse than the channel condition from $ \mathrm{V}_\alpha $ to BS, that is, $ R_{\beta \alpha} $ is less than  $ R_{\beta b} $. Therefore, with the power increase, the task delay decreases. However, when $ P_\beta $ is greater than a threshold which makes $ R_{\beta} = R_{\beta b} $, the task delay of ${\mathrm{V}_\beta}$ increases with the growth of $ P_\beta $.
   
   As for the results for  ${\mathrm{V}_\alpha}$, it can be seen from Fig. \ref{fig6} that $ D_\alpha $ in NOMA is larger than that in OMA modestly. Even if the task completion delay of ${\mathrm{V}_\alpha}$ in the NOMA scheme increases with the increase of power, ${\mathrm{V}_\alpha}$ still has a superior latency performance compared with ${\mathrm{V}_\beta}$. It can be concluded that our proposed GPM can significantly reduce the task completion delay of ${\mathrm{V}_\beta}$ at the extremely low cost of ${D_\alpha}$.

\setlength{\textfloatsep}{5pt}
	\begin{figure}[t]
	\centering
	\includegraphics[width=3.5in,height=2.5in]{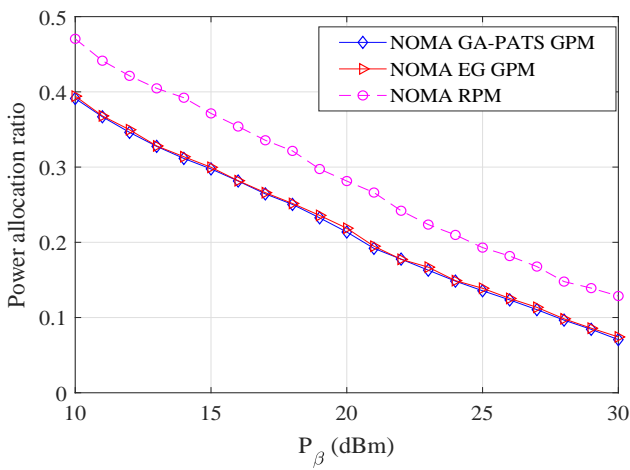}		
	\caption{The power allocation ratio $ \lambda $ of different schemes versus $ P_\beta $.}
	\label{fig7}
   \end{figure}
   Figure \ref{fig7} shows the relationships between the power allocation ratio $ \lambda $ and the power of $ \mathrm{V}_\beta $ in different NOMA schemes. It can be observed from Fig. \ref{fig7} that as the power increases, the power allocation ratio keeps decreasing. According to the analysis in Fig. \ref{fig6}, the completion delay of a single task depends on the minimum of  $ R_{\beta \alpha} $ and $ R_{\beta b} $. In order to minimize the task delay of $ V_\beta $ and ensure low-latency communication of $ \mathrm{V}_\alpha $, the parameters will be changed to make the values of  $ R_{\beta \alpha} $ and $ R_{\beta b} $ as equal as possible during optimization. Increasing power can rise $ R_{\beta \alpha} $, so that $ R_{\beta b} $  should also increase, which requires a decrease in  $ \lambda $.

	\begin{figure}[t]
	\setlength{\textfloatsep}{5pt}
	\centering	   
	\includegraphics[width=3.5in,height=2.5in]{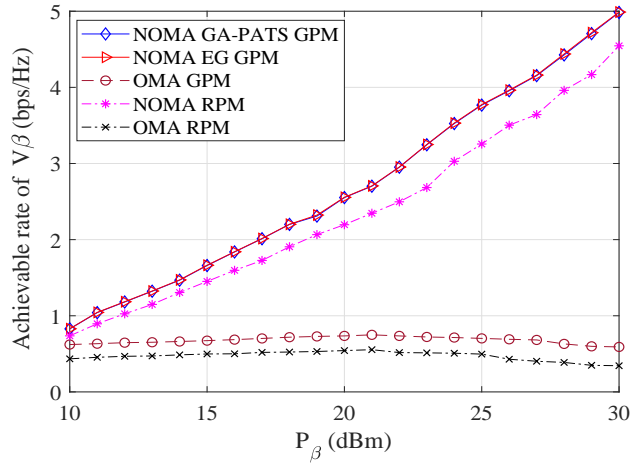}%scale=0.6			
	\caption{Achievable transmission rate of $ \mathrm{V}_\beta $ in different schemes versus $ P_\beta $.}
	\label{fig8}
    \end{figure}
    Figure \ref{fig8} reveals the achievable transmission rate of $ \mathrm{V}_\beta $ versus $ P_\beta $ for different schemes. It can be concluded from the figure that, compared with OMA, NOMA technology can bring significant performance gain in achievable transmission rate of $ \mathrm{V}_\beta $, and the larger the power of $ \mathrm{V}_\beta $, the more obvious the disparity between NOMA and OMA. On the other hand, this figure also verifies that the proposed vehicle GPM is superior to the RPM in terms of achievable transmission rate.
    Combining the results of Figure \ref{fig6}, \ref{fig7}, and \ref{fig8}, we can find that when the power is lager, the increase of power allocation ratio plays a major role in the reduction of the delay. This is why in Fig. \ref{fig6}, as the value of  $ P_\beta $ is greater than a threshold, the trend of achievable transmission rate in OMA is opposite to that in NOMA.

    \begin{figure}[t]
    	\setlength{\textfloatsep}{0pt}
    	\centering
    	\includegraphics[width=3.5in,height=2.5in]{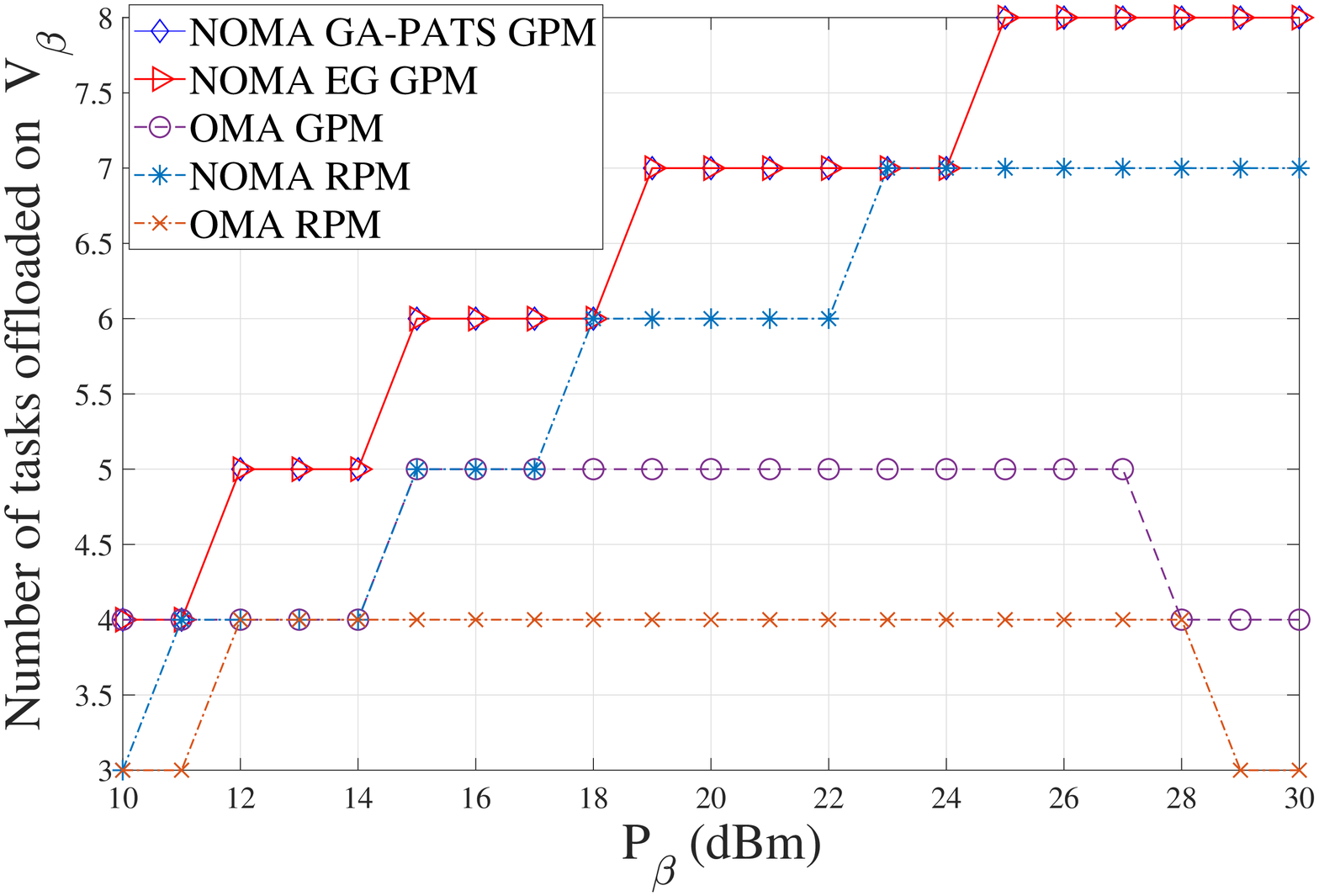}		
    	\caption{The number of tasks offloaded on $  V_\beta $ in different schemes versus $ P_\beta $.}
    	\label{fig9}
    \end{figure}
     Figure \ref{fig9}  illustrates the relationships between the number of tasks offloaded to MEC on $ \mathrm{V}_\beta $ and the $ P_\beta$. Combining Figure \ref{fig8}  and \ref{fig9}, it can be seen that as the transmission rate increases, the number of tasks offloaded to the MEC also increases, and the number of tasks performed locally will decrease. In summary, the transmission rate will affect the task division of $ \mathrm{V}_\beta $. It can also be concluded that compared with OMA scheme, NOMA can offload more tasks to MEC, while most of the tasks in OMA scheme are executed locally.

     It can be summarized from the above figures that our scheme empowered by the designed GPM and NOMA philosophy can significantly reduce the task completion delay and has better safety performance compared with OMA and RPM.
    
    \section{Conclusions}
    In this paper, we have studied a multi-vehicle multi-task NOMA-MEC system with a passive eavesdropping vehicle. Firstly, we have proposed the GPM to enhance the performance of edge vehicles. Then, AN has been utilized to combat the eavesdropper. On this basis, for each pair of vehicles, an approximate expression of SOPS has been obtained in closed form by deriving the SOP of center and edge vehicles sequentially. Additionally, to minimize the total task completion time of edge vehicles, we have designed the GA-PATS algorithm to settle the joint optimization problem of vehicle task division, power allocation, and transmit beamforming. Finally, the simulation results have demonstrated that compared with the benchmark schemes, the application of AN and vehicle GPM in the NOMA-MEC networks could enhance system security and reduce the task completion delay of edge vehicles.

\end{document}